\documentclass[12pt,a4paper,twoside]{article}
\usepackage{graphicx}   
\begin{document}

\centerline{\bf \Large Search for emission of unstable $^{8}$Be clusters}
\vskip 0.3cm 
\centerline{\bf \Large from hot $^{40}$Ca and $^{56}$Ni nuclei} 

\vskip 0.9cm 

\noindent
M. Rousseau$^{a}$, C. Beck$^{a}$, C. Bhattacharya$^{a,e}$, V. Rauch$^{a}$, 
O. Dorvaux$^{a}$, K. Eddahbi$^{a}$, C. Enaux$^{a}$, R.M. Freeman$^{a}$, 
F. Haas$^{a}$, A. Hachem$^{c}$, D. Mahboub$^{f}$, E. Martin$^{c}$, 
P. Papka$^{a}$, S.J. Sanders$^{d}$, O. Stezowski$^{g}$, A. 
Szanto de Toledo$^{b}$, and S. Szilner$^{a}$

\vskip 0.5cm

\leftline{\it a) Institut de Recherches Subatomiques, F-67037 Strasbourg,
Cedex 2, France } 

\leftline{\it b) Instituto de F\'{i}sica da Universidade de S\~ao Paulo, S\~ao
Paulo, Brazil} 

\leftline{\it c) Universit\'e de Nice-Sophia-Antipolis, Nice, France}

\leftline{\it d) University of Kansas, Lawrence, KS 66045, USA}

\leftline{\it e) Present address : VECC Calcutta, India}

\leftline{\it f) University of Surrey, Guildford, UK}

\leftline{\it g) Institut de Physique Nucl\'eaire de Lyon, Lyon, France}

\vskip .5cm

{\bf ABSTRACT :} {\it The possible occurence of highly deformed configurations
is investigated in the $^{40}$Ca and $^{56}$Ni di-nuclear systems as formed in
the $^{28}$Si + $^{12}$C and $^{28}$Si + $^{28}$Si reactions, respectively, by
using the properties of emitted light charged particles. Inclusive as well as
exclusive data of the heavy fragments (A $\geq$ 6) and their associated light
charged particles (p, d, t, and $\alpha$-particles) have been collected at the
IReS Strasbourg {\sc VIVITRON} Tandem facility with two bombarding energies
$E_{lab}$($^{28}$Si) = 112 and 180 MeV by using the {\sc ICARE} charged
particle multidetector array, which consists of nearly 40 telescopes. The
measured energy spectra, velocity distributions, in-plane and out-of-plane
angular correlations are analysed by Monte Carlo {\sc CASCADE}
statistical-model calculations using a consistent set of parameters with
spin-dependent level densities. Although significant deformation effects at
high spin are needed, the remaining disagreement observed in the $^{28}$Si +
$^{12}$C reaction for the S evaporation residue suggests an unexpected large
unstable $^{8}$Be cluster emission of a binary nature.} 

\vskip 0.8cm

\centerline {\bf I. INTRODUCTION}

\vskip 0.8cm

The formation and decay processes of {\bf light} di-nuclear systems (in the
A$_{CN}$ $\leq$ 60 mass region), produced by low-energy (E$_{lab}$ $\leq$ 7
MeV/nucleon) heavy-ion reactions, has been studied for a long time both from
the experimental and the theoretical points of view [1]. In most of the
reactions studied, whereas the general conclusions about the formation
probability for the compound nucleus (CN) and the characteristic features of
its decay could clearly been drawn, the properties of the observed fully energy
damped yields are still debated in terms of either a fusion-fission (FF)
mechanism [1,2,3], which may be considered as the emission of complex (or
intermediate mass) fragments, or a deep-inelastic (DI) orbiting [4] mechanism
behavior, the latter being found to be particularly competitive in the
$^{28}$Si+$^{12}$C reaction [5]. Since many of the conjectured features for
orbiting yields are similar to that expected for the FF mechanism, it is
difficult to fully discount FF as a possible explanation for the large energy
damped $^{28}$Si+$^{12}$C yields [4,5] and, thus, FF, DI orbiting, and even
molecular-resonance behavior may coexist in the large-angle yields of the
$^{28}$Si+$^{12}$C reaction [1]. 

Superdeformed (SD) rotational bands have been found in various mass regions (A
= 60, 80, 130, 150 and 190) and, very recently, one SD band has also been
discovered in the N = Z nucleus $^{36}$Ar [6]. This new result makes that the A
$\approx$ 40 mass region is of special interest for the nuclear $\gamma$-ray
spectroscopists. With this respect the study of the N = Z nucleus $^{40}$Ca is
relevant. Since the detection of light charged particles (LCP) is relatively
simple, the analysis of their spectral shapes is another good tool in exploring
nuclear deformation and other properties of hot rotating nuclei at high angular
momenta. Experimental evidence for angular momentum dependent spectral shapes
has already been widely discussed in the literature [7-11]. Thus we decided to
investigate the $^{40}$Ca nucleus produced through the $^{28}$Si + $^{12}$C
reaction at the following bombarding energies E$_{lab}$($^{28}$Si) = 112 and
180 MeV. Data have also been collected for the $^{28}$Si + $^{28}$Si reaction
(leading to the N = Z nucleus $^{56}$Ni) in the same experimental conditions. 

\vskip 0.5 cm

\centerline {\bf II. EXPERIMENTAL RESULTS  }

\vskip 0.5 cm

\centerline {\bf A. Experimental techniques : }

\vskip 0.3cm

The experiments were performed at the IReS Strasbourg VIVITRON Tandem facility
using 112 MeV and 180 MeV $^{28}$Si beams which were incident on $^{12}$C (160
and 180 $\mu$g/cm${^2}$ thick) and $^{28}$Si (180 and 230 $\mu$g/cm${^2}$
thick) targets mounted in the ICARE scattering chamber [12]. Both
the heavy ions (A $\geq$ 6) and their associated LCP's (p, d, t, and $\alpha$)
were detected using the {\bf ICARE} charged particle multidetector
array [12] which consists in nearly 40 telescopes in coincidence.
The heavy fragments, i.e. ER, quasi-elastic (QE), DI, and FF fragments, were
detected in 10 heavy-ion telescopes, each consisting of an ionisation chamber
(IC) followed by a silicon surface-barrier diode of 500 $\mu$m effective
thickness. Both the in-plane and out-of-plane coincident LCP's were detected
using either 3 three-elements light-ion telescopes (Si 40 $\mu$m, Si 300
$\mu$m, CsI(Tl) 2 cm) or 24 two-elements light-ion telescopes (Si 40 $\mu$m, 
CsI(Tl) 2 cm) and two light-ion telescopes (IC, Si 500 $\mu$m) located at the
most backward angles. The CsI(Tl) scintillators were coupled to photodiode
readouts. The IC's were filled with isobutane and the pressures were kept at 30
torr and at 60 torr for detecting heavy fragments and light fragments,
respectively. The acceptance of each telescope was defined by thick aluminium
collimators. The distances of these telescopes from the target ranged from 10.0
to 30.0 cm, and the solid angles varied from 1.0 msr at the most forward
angles to 5.0 msr at the backward angles. 

The different energy calibrations of the {\sc ICARE} multidetector array have
been done using radioactive sources in the range of 5 to 9 MeV, a precision
pulser, elastic scattering yields from $^{28}$Si on $^{197}$SAu, $^{28}$Si and
$^{12}$C targets, and data from the $^{16}$O+$^{12}$C reaction measured at
E$_{lab}$ = 53 MeV [12]. In this last case, $\alpha$-particles emitted in the
$^{12}$C($^{16}$O,$\alpha$)$^{24}$Mg$^{*}$ reaction provide known energies from
$^{24}$Mg excited states and allow for the $\alpha$-particle calibration of the
backward angle detectors. The proton calibration was done with scattered
protons from Formvar targets in reverse kinematics reactions with both
$^{28}$Si and $^{16}$O beams. 

\vskip 0.5 cm

\centerline {\bf B. Experimental data : }

\vskip 0.3cm

The inclusive cross sections for C, N and O fragments produced in the
$^{28}$Si+$^{12}$C reaction [13] have been found to be in good
agreement with the previously measured excitation functions [5].
Typical inclusive energy spectra of $\alpha$-particles are shown in Fig.~1 at
the indicated angles for the $^{28}$Si+$^{28}$Si reaction at
E$_{lab}$($^{28}$Si) = 180 MeV. \\ 

        \begin{figure}[ht!]
        \begin{center}
        \includegraphics[width=12cm,clip=true,draft=false]        
        {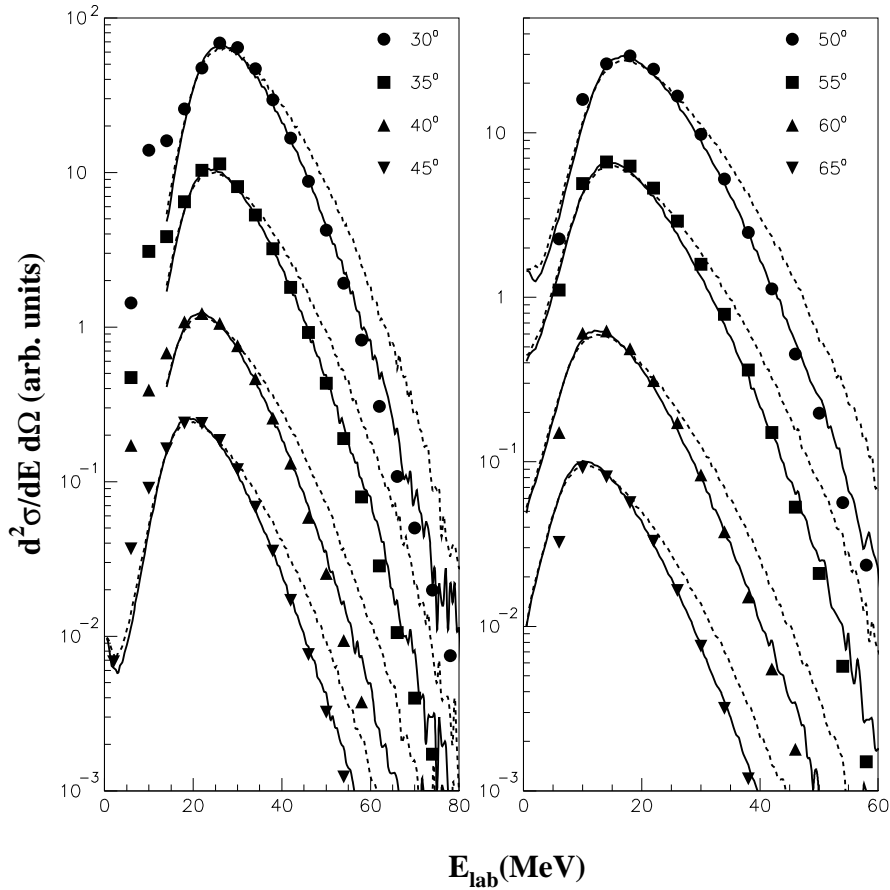}
        \parbox{140mm}{\caption
                      {\label{figure1}  
                      {\it  Energy spectra of $\alpha$-particles measured in a single mode for
the $^{28}$Si(E$_{lab}$ = 180 MeV) + $^{28}$Si reaction between
$\theta$$_{lab}$ = 30$^\circ$ and 65$^\circ$. The solid and dashed lines are
statistical-model calculations discussed in the text.}}} 
        \end{center}
        \end{figure}

The LCP energy spectra have Maxwellian shapes with an exponential fall-off at
high energy (their shape and high-energy slopes are essentially independent of
c.m. angle) which reflects a relatively low temperature of the decaying
nucleus. This is the signature of a statistical de-excitation process arising
from a thermalized source such as the $^{56}$Ni CN. Similar LCP spectra have
been measured for the lowest bombarding energy. In each case, The sizeable
low-energy backgrounds, disappearing in coincidence with ER's (see Fig.~4),
arise primarly from binary reaction mechanisms. We have observed the same
behavior in the $^{28}$Si+$^{12}$C reaction for both the $\alpha$-particles,
and the protons at the two incident energies E$_{lab}$ = 112 and 180 MeV, in
agreement with a previous study at E$_{lab}$ = 150 MeV [14], but in
strong contrast to previous published data of proton energy spectra that reveal
at least a second unexpected component [15]. 

The velocity contour maps of the invariant cross sections provide an overall
picture to characterize the emitting sources with respect to the reaction
mechanism. Therefore, to have a better insight into the nature of these
emitters, the measured energy spectra of the LCP's have been transformed into
invariant cross-section plots in the velocity space. Fig.~2 shows such typical
plots of invariant cross-section in the (V$_{\parallel}$,V$_{\perp}$) plane for
$\alpha$-particles and protons, respectively ; the axes V$_{\parallel}$ and
V$_{\perp}$ denote laboratory velocity components parallel and perpendicular to
the beam, respectively. The circles are defined to visualize the maxima of
intensity production. All spectra can be understood by assuming an evaporative
process from a single source, i.e., the thermally equilibrated $^{40}$Ca and
$^{56}$Ni CN's. \\

        \begin{figure}[ht!]
        \begin{center}
        \includegraphics[width=11cm,clip=true,draft=false]        
        {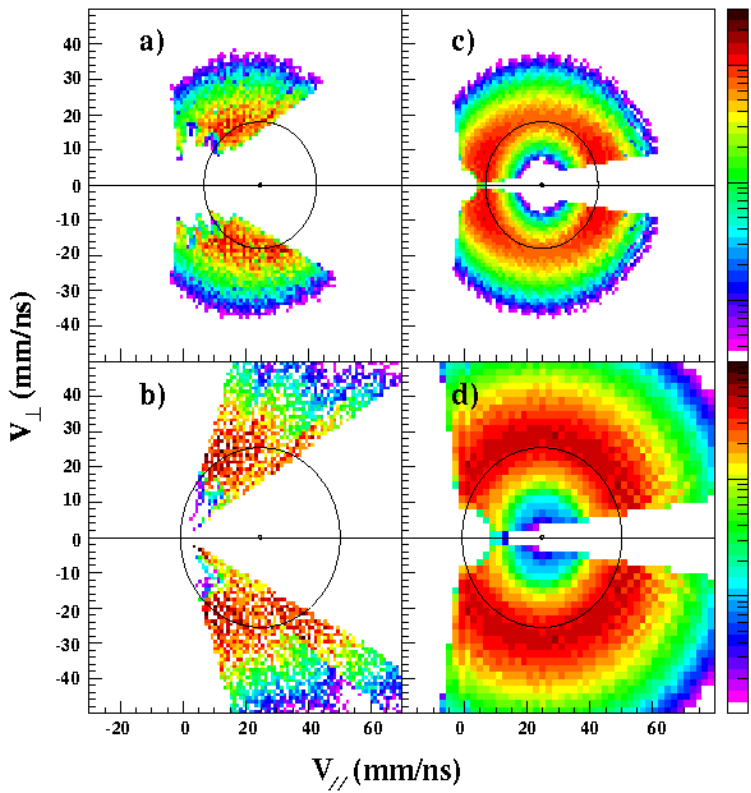}
        \parbox{140mm}{\caption
                      {\label{figure2}  
                      {\it  Inclusive $\alpha$-particle (a) and proton (b) invariant cross
sections measured in the (V$_{\perp}$,V$_{\parallel}$) plane for the 180 MeV
$^{28}$Si+$^{12}$C reaction. (c) and (d) are statistical-model calculations
discussed in the text.}}} 
        \end{center}
        \end{figure}

It is clear from this figure that the invariant cross section contours fall on
circles centered at V$_{CN}$, as expected for a complete fusion-evaporation
(CF) mechanism. 

The in-plane angular correlation of $\alpha$-particles in coincidence with all
the ER's, produced in the 180 MeV $^{28}$Si+$^{12}$C reaction, is shown in
Fig.~3. The angular correlation is strongly peaked at the opposite side of the
ER detector located at $\theta^{ER}_{lab}$ = -10$^{\circ}$ with respect to the
beam. This peaking on the opposite side of the beam is due to the momentum
conservation. The solid lines shown in the figure are the results of
statistical-model predictions for CF and equilibrium decay using the code {\sc
CACARIZO}, to be discussed in the following Section. It is seen that the
shape of the experimental angular correlations is well reproduced by the
theory. However, the excess of yields observed at backward angles
($\theta_{lab}$= +50$^\circ$ to +90$^\circ$) indicates the occurence of a
non-evaporative process possibly of a binary nature. 

        \begin{figure}[ht!]
        \begin{center}
        \includegraphics[width=12cm,clip=true,draft=false]        
        {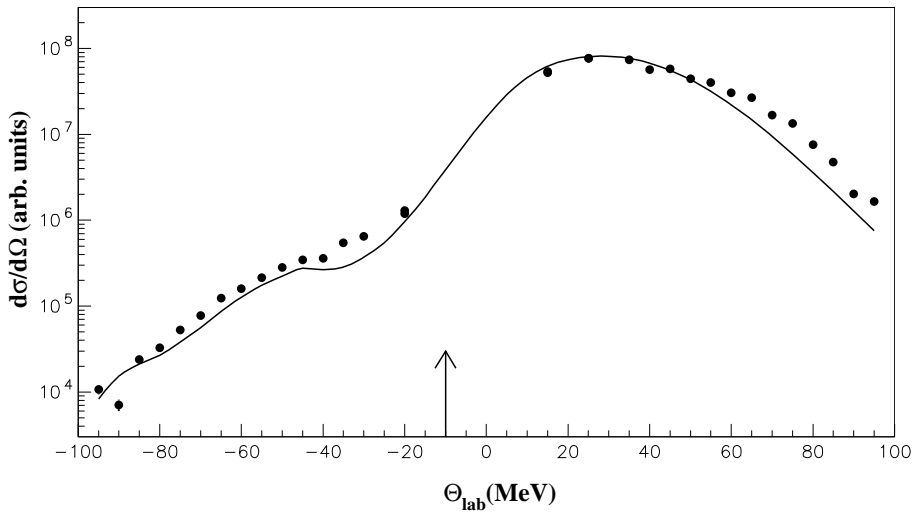}
        \parbox{140mm}{\caption
                      {\label{figure3}  
                      {\it  In-plane 180 MeV $^{28}$Si+$^{12}$C angular correlations of
coincident $\alpha$-particles. The points correspond to the data, and the solid
line to statistical-model calculations discussed in the text. The arrow
indicates the position of the IC detector at $\theta_{lab}$= -10$^\circ$.}}} 
        \end{center}
        \end{figure}

The spectral shapes of the coincident $\alpha$-particles, displayed in Fig.~5
for the 180 MeV $^{28}$Si+$^{12}$C reaction are also analysed using the
statistical-model code {\sc CACARIZO} [7,17], as described in
the next Section. 

        \begin{figure}[ht!]
        \begin{center}
        \includegraphics[width=12cm,clip=true,draft=false]        
        {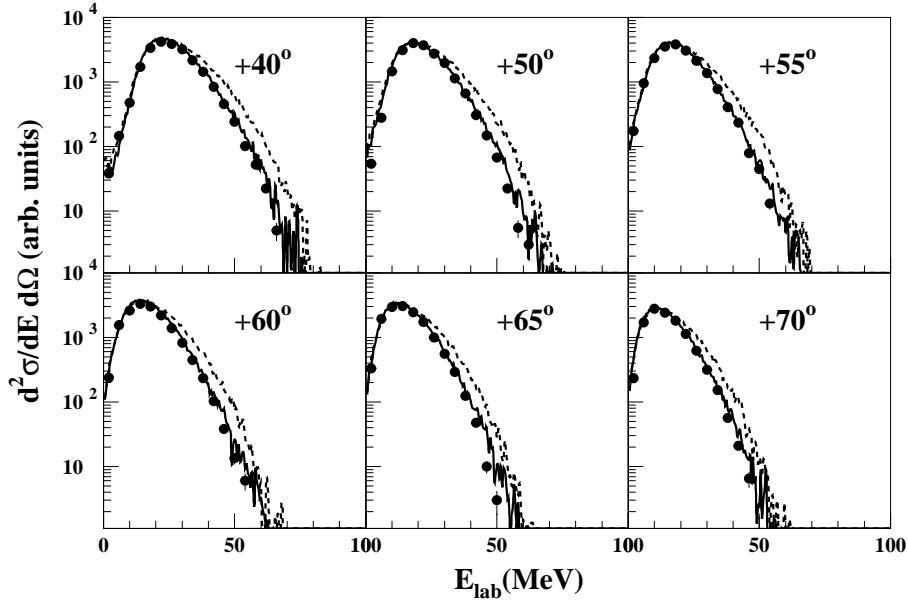}
        \parbox{140mm}{\caption
                      {\label{figure4}  
                      {\it  Exclusive energy spectra of $\alpha$-particles emitted in
coincidence with all ER's detected at -10$^\circ$, at the indicated laboratory
angles, in the 180 MeV $^{28}$Si+$^{28}$Si reaction. The solid and dashed lines
are statistical-model calculations discussed in the text.}}} 
        \end{center}
        \end{figure}

\vskip 0.5cm

\centerline {\bf III. DISCUSSION }

\vskip 0.3cm

A Hauser-Feshbach [1] analysis of the data has been performed
using {\sc CACARIZO} [7,17], the Monte Carlo version of the
statistical-model code {\sc CASCADE}, in which the experimental geometry of the
{\sc ICARE} detectors is taken into account. The angular momentum distributions
needed as the primary input for the calculations were taken from compiled
$^{28}$Si+$^{12}$C [16] and $^{28}$Si+$^{28}$Si [18]
CF data, respectively. The other standard ingredients for statistical-model
calculations, namely the nuclear level densities and the barrier transmission
probabilities, are obtained from studies of the evaporated LCP spectra. In
recent years, it has been observed that the standard statistical model cannot
predict satisfactorily the shape of the evaporated $\alpha$-particle energy
spectra [7-11], and the calculated
average energies of the $\alpha$-particles are found to be much higher than the
corresponding experimental results. Several attempts have been made to explain
this anomaly either by changing the emission barrier or by using spin-dependent
level densities. Adjusting the emission barriers and corresponding transmission
probabilities affect the lower energy part of the calculated evaporation
spectra. On the other hand the high-energy part of the spectra depends
crucially on the available phase space obtained from the level densities at
high spin. In hot rotating nuclei formed in heavy-ion reactions, the energy
level density at higher angular momentum is spin dependent. The level density,
$\rho(E,J)$, for a given angular momentum $J$ and energy $E$ is given by the
well known Fermi gas expression : 

\begin{equation}
\rho(E,J) = {\frac{(2J+1)}{12}}a^{1/2}
           ({\frac{ \hbar^2}{2 {\cal J}_{eff}}}) ^{3/2}
           {\frac{1}{(E-\Delta-t-E_J)^2} }exp(2[a(E-\Delta-t-E_J)]^{1/2})
\label{lev}
\end{equation}

where $a$ is the level density parameter, t is the ``nuclear" temperature and
$\Delta$ is the pairing correction, E$_J$ = $\frac{ \hbar^2}{2 {\cal
J}_{eff}}$J(J+1) is the rotational energy, ${\cal J}_{eff}= {\cal J}_0 \times
(1+\delta_1J^2+\delta_2J^4)$ is the effective moment of inertia,  ${\cal J}_0$
is the rigid body moment of inertia and $\delta_1$ and $\delta_2$ are
deformability parameters [7,8,10,11]. By changing the deformability parameters
$\delta_1$ and $\delta_2$ one can simulate the spin-dependent level density
[7,8]. The {\sc CACARIZO} calculations have been performed using two sets of
input parameters : the first one with standard rotating liquid drop model
(RLDM) [1] parameters, and the second one with non-zero values for the
deformability parameters [13,19,20,21] which are listed in Table I. 

\hspace{-2cm}
\renewcommand{\baselinestretch}{1.2}
\begin{table}[h!]
\hspace{-2cm}
\begin{tabular}{|c|c|c|c|c|c|c|c|c|}
\hline
Reaction & C.N. & Energy [MeV]& $l_{cr}$ [$\hbar$] & $\delta_1$ & $\delta_2$ & b/a & $\beta_2$ & Reference\\
\hline
$^{28}$Si+$^{28}$Si & $^{56}$Ni & 111.6 and 180 & 34 and 37 & 1.2$\cdot$10$^{-4}$ & 1.1$\cdot$10$^{-7}$ & 1.6 and 1.7 & .49 and .50 & This work
\\
\hline
$^{28}$Si+$^{12}$C & $^{40}$Ca  & 111.6 and 180 & 21 and 27 & 2.5$\cdot$10$^{-4}$ & 5.0$\cdot$10$^{-7}$ & 1.7 and 1.8 & .47 and .51 & This work
\\
\hline
$^{28}$Si+$^{12}$C & $^{40}$Ca  &  150 & 26 & 6.5$\cdot$10$^{-4}$ & 3.3$\cdot$10$^{-7}$ & 1.7 & .51 &[5] \\
\hline
$^{35}$Cl+$^{24}$Mg & $^{59}$Cu  & 260 & 37 & 1.1$\cdot$10$^{-4}$ & 1.3$\cdot$10$^{-7}$ & 1.4 to 2.0 & .46 to .53 & [22]\\
\hline
$^{32}$S+$^{27}$Al & $^{59}$Cu  & 100 to 150 & 27 to 39 & 2.3$\cdot$10$^{-4}$ & 1.6$\cdot$10$^{-7}$ & 2.0 & .53 & [7] \\
\hline
$^{32}$S+$^{27}$Al & $^{59}$Cu  &  100 to 150 & 27 to 42 & 1.3$\cdot$10$^{-4}$ & 1.2$\cdot$10$^{-7}$ & 1.5 to 2.2 & .48 to .54 & [10] \\
\hline
$^{28}$Si+$^{27}$Al & $^{55}$Co  &  150 & 42 & 1.8$\cdot$10$^{-4}$ & 1.8$\cdot$10$^{-7}$ & 1.3 & .46 & [11] \\
\hline
\end{tabular}
\renewcommand{\baselinestretch}{1.}
\caption{\label{comp-cacar}\textit{Input parameters of the CACARIZO calculations.}}
\end{table}

For the 180 MeV $^{28}$Si + $^{28}$Si reaction, the shape of the inclusive (see
Fig.~1) and exclusive (see Fig.~4) $\alpha$ energy spectra are well reproduced
by using deformation
effects [13,19,20,21]. The dashed
lines in Figs.~1 and 4 show the predictions of {\sc CACARIZO} using the
standard RLDM deformation parameter set with no extra deformation ($\delta _1$
= $\delta_2$ = 0). It is clear that the average energies of the measured
$\alpha$ energy spectra are lower than those predicted by the standard
statistical-model calculations. The solid lines show the predictions of {\sc
CACARIZO} using the parameter set with $\delta_1$ = 1.2 x 10$^{-4}$ and
$\delta_2$ = 1.1 x 10$^{-7}$ chosen to reproduce the data consistently at the
two bombarding energies . The shapes of the inclusive as well as the exclusive
$\alpha$ energy spectra are well described after including these significative
deformation effects. {\sc CACARIZO} calculations of the invariant cross sections, plotted
in Fig.~2, reproduce well the data obtained in the 180 MeV $^{28}$Si+$^{12}$C
reaction for both $\alpha$-particles and protons, respectively. In this case the {\sc CACARIZO} parameters are similar to
those employed in a previous study of the 130 MeV $^{16}$O+$^{24}$Mg
reaction [17], with the use of the angular momentum dependent level
densities. The exclusive
energy spectra of $\alpha$-particle measured in coincidence with individual S
and P ER's, which are shown in Fig.~5 for E$_{lab}$ = 180 MeV, are quite
interesting. The energy spectra associated with S are completely different from
those associated with P [13]. The latter are reasonably well
reproduced by the CACARIZO curves whereas the model could not predict the shape
of the spectra obtained in coincidence with S at backward angle ($\theta_{\alpha}\geq 70^{\circ}$). Additional non-statistical
components appear to be significant in this case. This is consistent with the
discrepancies also observed at backward angles in the in-plane angular
correlations of Fig.~3. The same observations can be made at the lower
bombarding energy E$_{lab}$ = 112
MeV [19-21]. We will show that the
hypothesis of a contributions arising from the decay of unbound $^{8}$Be clusters
produced in a binary reaction $^{40}$Ca $\rightarrow$ $^{32}$S+$^{8}$Be is
consistent with the experimental results. 
        \begin{figure}[ht!]
        \begin{center}
        \includegraphics[width=11cm,clip=true,draft=false]        
        {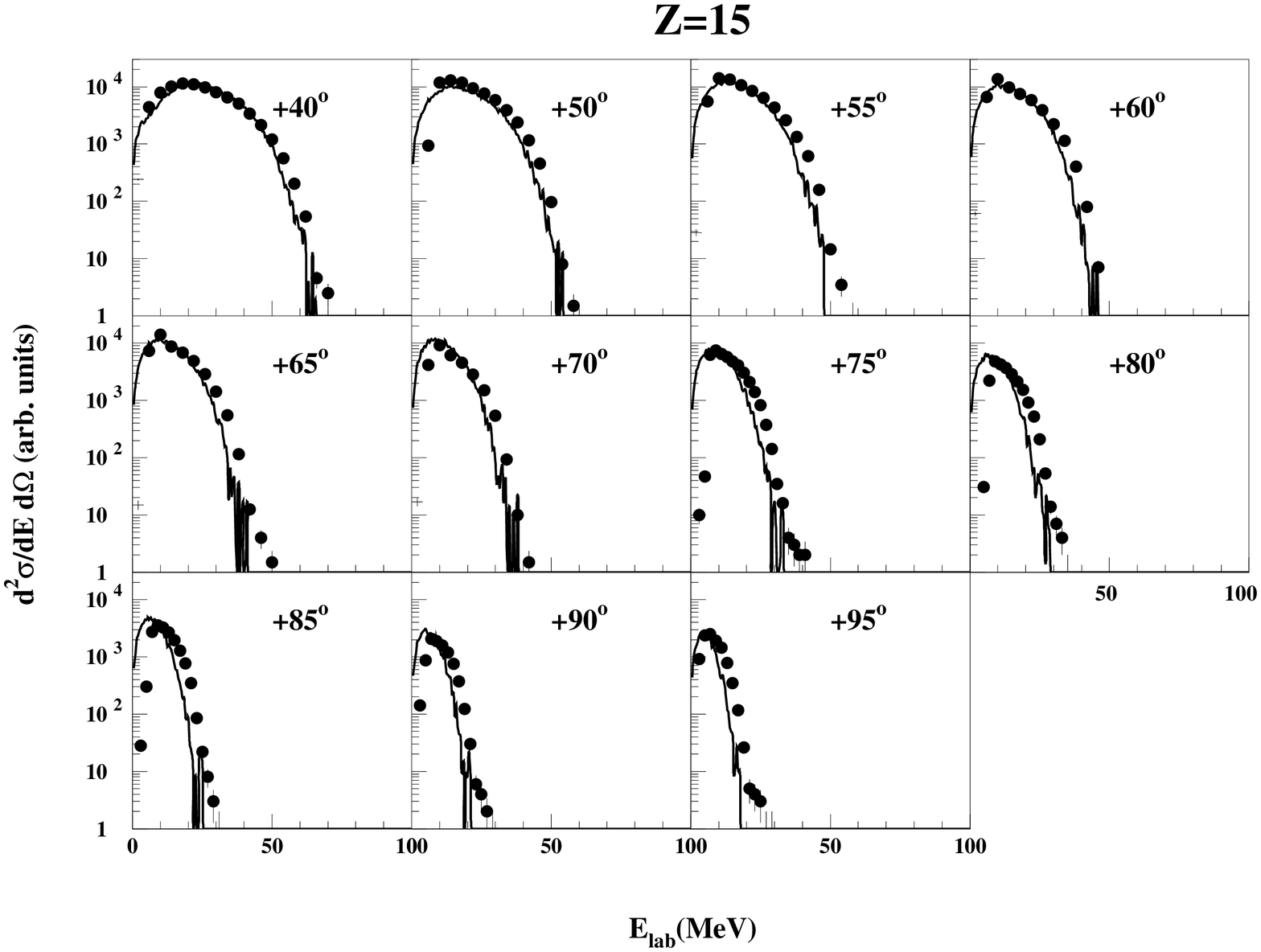}
        \includegraphics[width=11cm,clip=true,draft=false]        
        {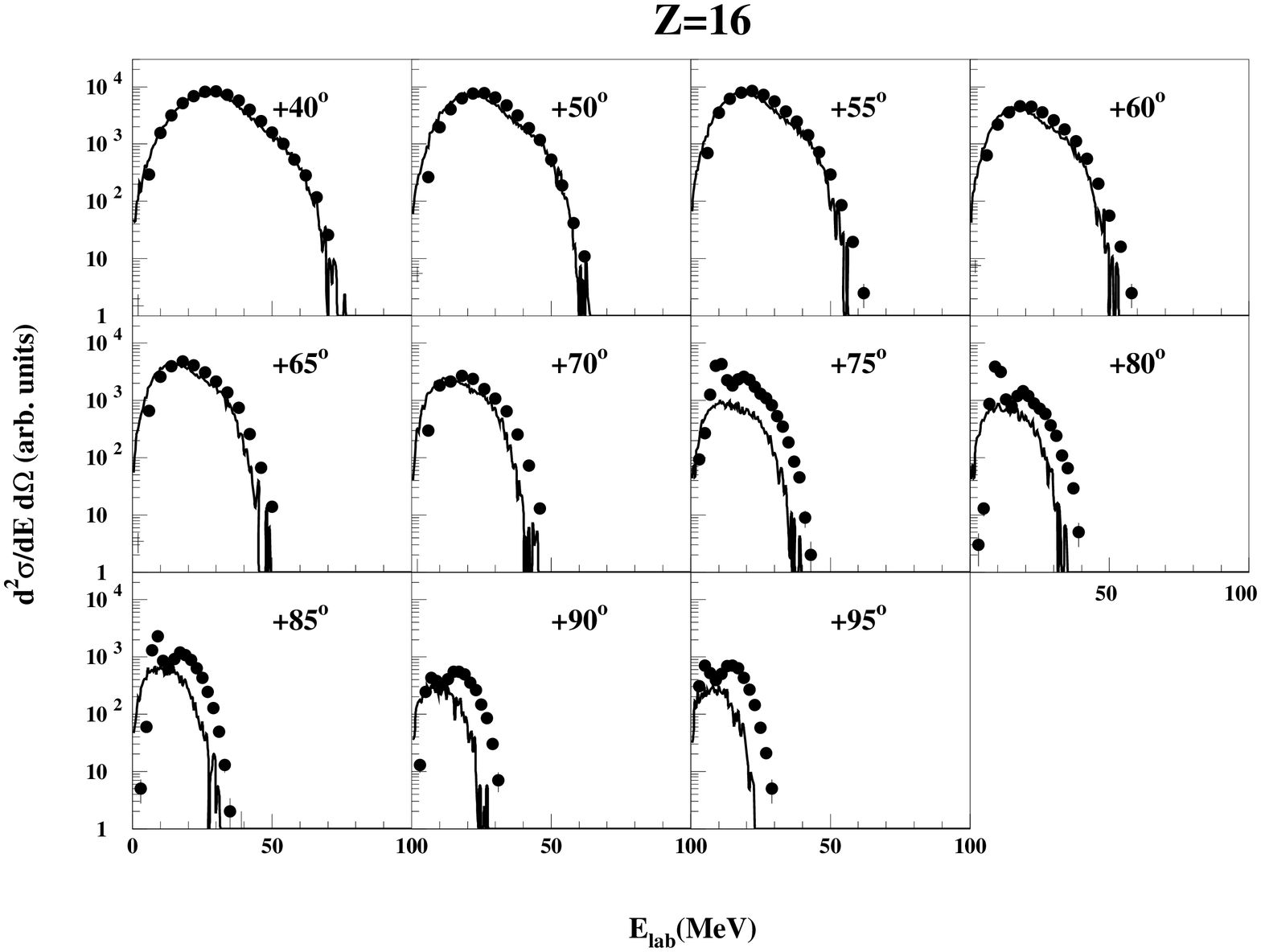}
        \parbox{140mm}{\caption
                      {\label{figure5}  
                      {\it  Exclusive energy spectra of $\alpha$-particles emitted in
coincidence with individual P and S ER's detected at -10$^\circ$, at the indicated laboratory
angles, in the 180 MeV $^{28}$Si+$^{12}$C reaction. The solid lines
are statistical-model calculations discussed in the text.}}} 
        \end{center}
        \end{figure}

        \begin{figure}[ht!]
        \begin{center}
        \includegraphics[width=11cm,clip=true,draft=false]        
        {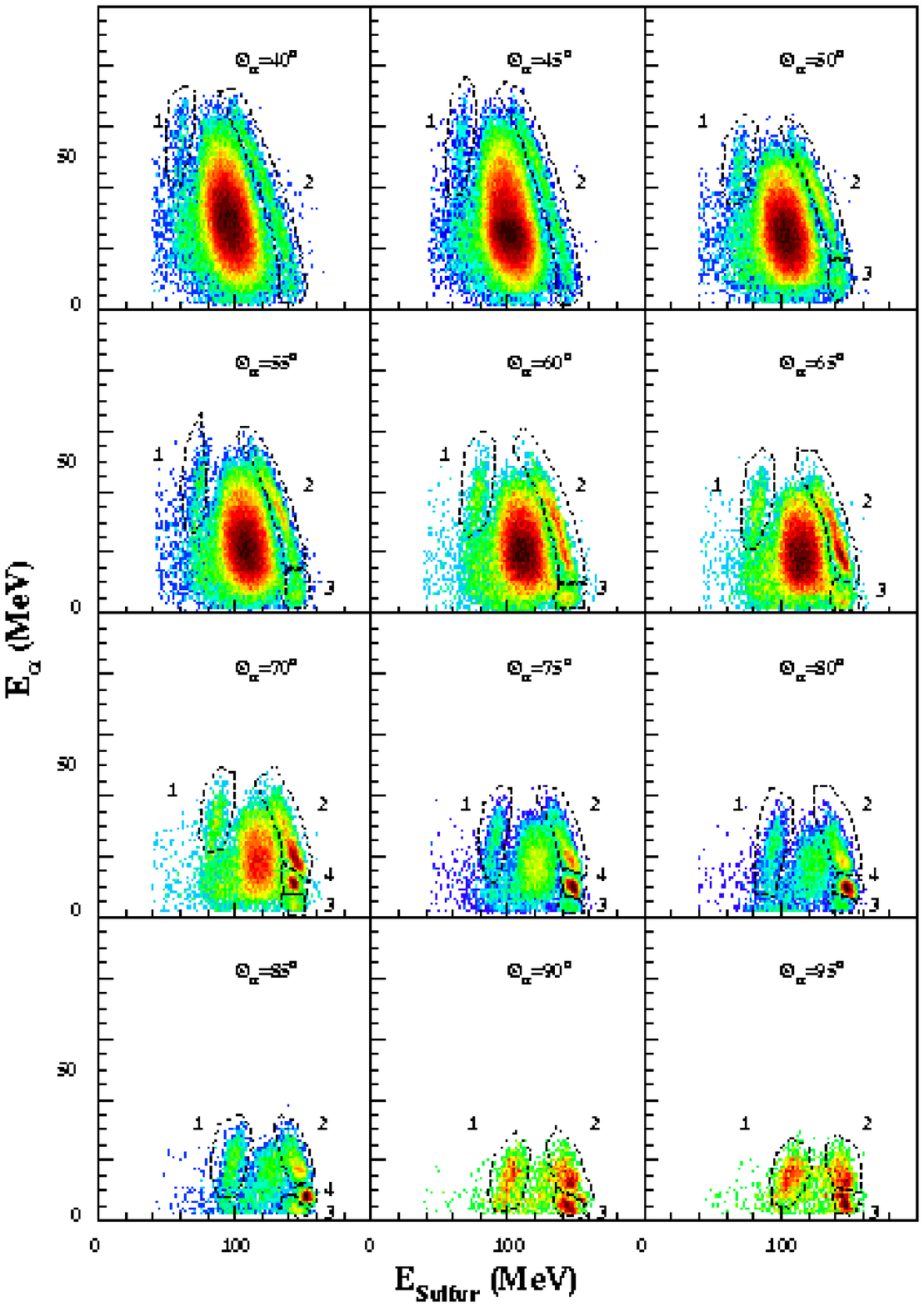}
        \parbox{140mm}{\caption
                      {\label{figure6}  
                      {\it  Energy-correlation plots between coincident $\alpha$-particles and S
ER's produced in the 180 MeV $^{28}$Si+$^{12}$C reaction. The heavy fragments
is detected at $\theta_{S}$ = -10$^\circ$ and $\alpha$'s angles are given in
the figure. The dashed lines correspond to different contours and their
associated labelling are discussed in the text.}}} 
        \end{center}
        \end{figure}

In Fig.~6 the energies of the $\alpha$-particles (detected in coincidence with
the S residues at the indicated angles) are plotted against the energies of the
S residues detected at $\theta_{S}$ = -10$^\circ$. At $\theta_{\alpha}$ =
+40$^\circ$, +45$^\circ$ and +50$^\circ$ the main bulk of events are from a statistical
origin, and consistent with CACARIZO calculations (see Fig.~7). For larger
angles, the two branches, labelled 1 and 2, are outside the "statistical
evaporation region", but still correspond to a evaporation process as shown by
the CACARIZO calculations displayed in Fig.~7. This two branches 1 and 2 corresponds to a 2-$\alpha$ channel with both $\alpha$-particles emitted respectively at backward and forward angles in the center of mass. 

        \begin{figure}[ht!]
        \begin{center}
        \includegraphics[width=11cm,clip=true,draft=false]        
        {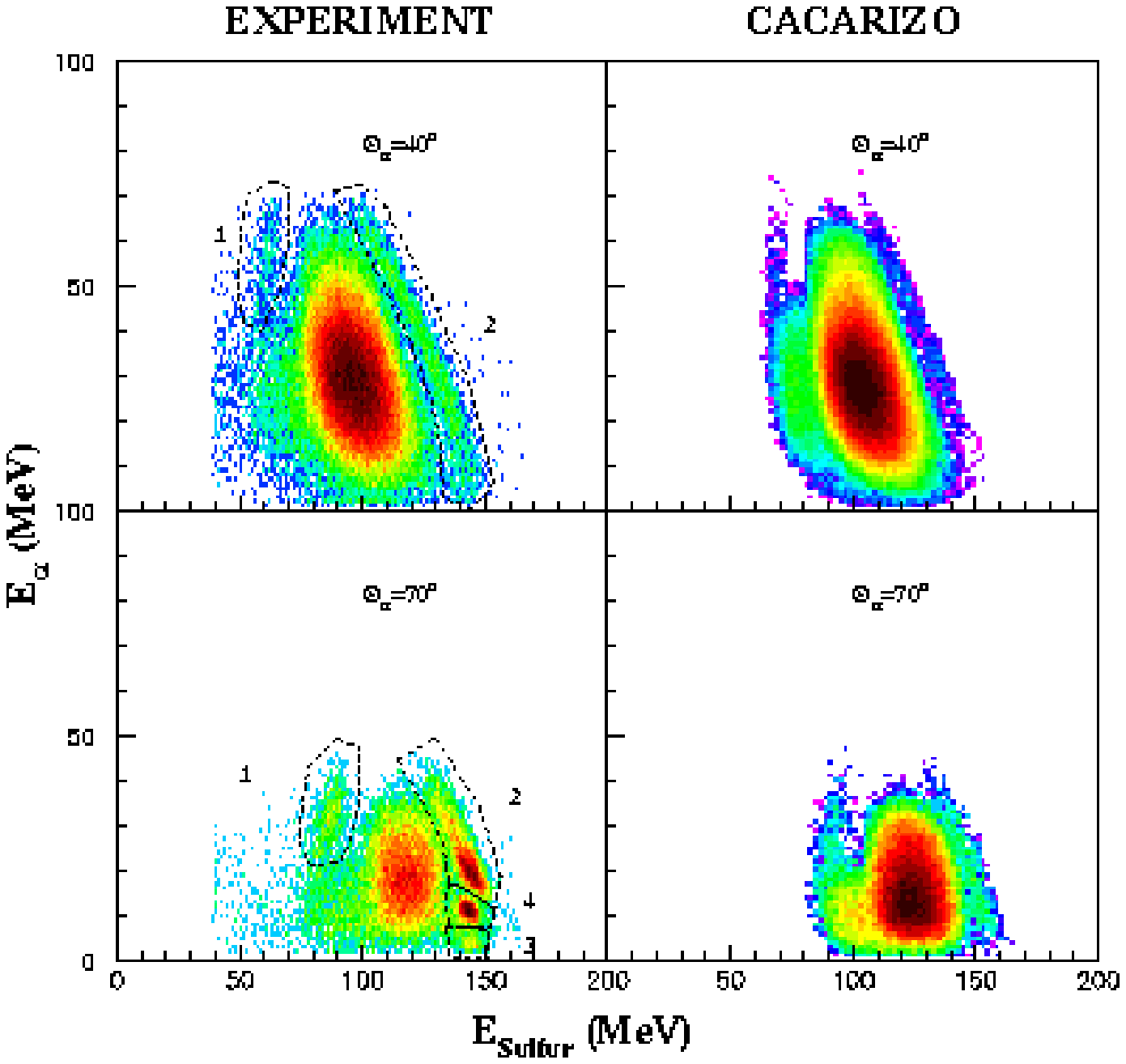}
        \parbox{140mm}{\caption
                      {\label{figure7}  
                      {\it  Experimental and calculated energy-correlation plots between
coincident $\alpha$-particles and S ER's produced in the 180 MeV
$^{28}$Si+$^{12}$C reaction. The S is detected at $\theta_{S}$ = -10$^\circ$
and $\alpha$'s at $\theta_{\alpha}$ = + 40$^\circ$,and + 70$^\circ$,
respectively. CACARIZO calculations are discussed in the text.}}} 
        \end{center}
        \end{figure}

However, at more backward angles other contributions, labelled 3 and 4, appear
more and more significantly. The observed "folding angles" are compatible with
the two-body kinematics required for the $^{32}$S+$^{8}$Be binary exit-channel.
On the other hand, the energy correlations for the $\alpha$-particles in
coincidence with the P residues (not shown) do not exhibit any of the two-body
branches and, thus, the "statistical evaporation region" alone, which is
consistent with the CACARIZO predictions at all the measured angles, appears
significantly. In the left side of Fig.~8 the excitation energy of $^{8}$Be,
calculated by assuming a two-body process, is presented for the contributions
labelled 2, 3, and 4 (Fig.~6) at different indicated angles. From 55$^\circ$ to
95$^\circ$ the main bulk of the yields is centered at around 3.06 $\pm$ 1.50
MeV, the energy of the first 2$^+$ excited level of $^{8}$Be; whereas, from
70$^\circ$ to 90$^\circ$ a second component is centered at the energy of the
ground state of $^{8}$Be (squared part). This last component corresponds to the
contribution 4 (Fig.~6). Fusion-fission calculations using the Extended
Hauser-Feshbach Method [3] fail to reproduce both the excitation energies of
the S residue, and the yields from the 3 and 4 contributions [12]. However,
these contributions can be interpreted as due to an incomplete fusion (ICF)
process governed by a $\alpha$-transfer reaction mechanism $^{28}$Si+$^{12}$C
$\rightarrow$ $^{32}$S$^{*}$+$^{8}$Be as proposed by Morgenstern et {\it et
al.} [23]. This conclusion is in agreement with previous inclusive results
published in Ref. [24]. In the cluster-transfer picture the reaction is
characterized by a "Q-value" window" centered at the so-called "Q-optimum",
which value can be estimated semi-classically by Q$_{opt}$ =
(Z$_{3}$Z$_{4}$/Z$_{1}$Z$_{2}$-1)E$_i^{c.m.}$, where the indices 1,2 and 3,4
indicate the entrance (i) and exit channel, respectively. The corresponding
excitation energy E$^{*}$ = Q$_{gg}$ - Q$_{opt}$, with Q$_{gg}$ is the
ground-state Q-value of the reaction. In this case the expected excitation
energy in the $^{32}$S nuclei is equal to 12.9 MeV. The right hand side of
Fig.~8 represents the calculated excitation energy of $^{32}$S in coincidence
with the ground state (g.s), and with the first excited state of ${^8}$Be,
respectively. The dashed lines correspond to E$*{*}$ = 12.9 MeV, the energy
expected for a $\alpha$-transfer reaction mechanism. In both cases the
excitation energies of $^{32}$S are consistent with these values. 

        \begin{figure}[ht!]
        \begin{center}
        \includegraphics[width=16cm,clip=true,draft=false]        
        {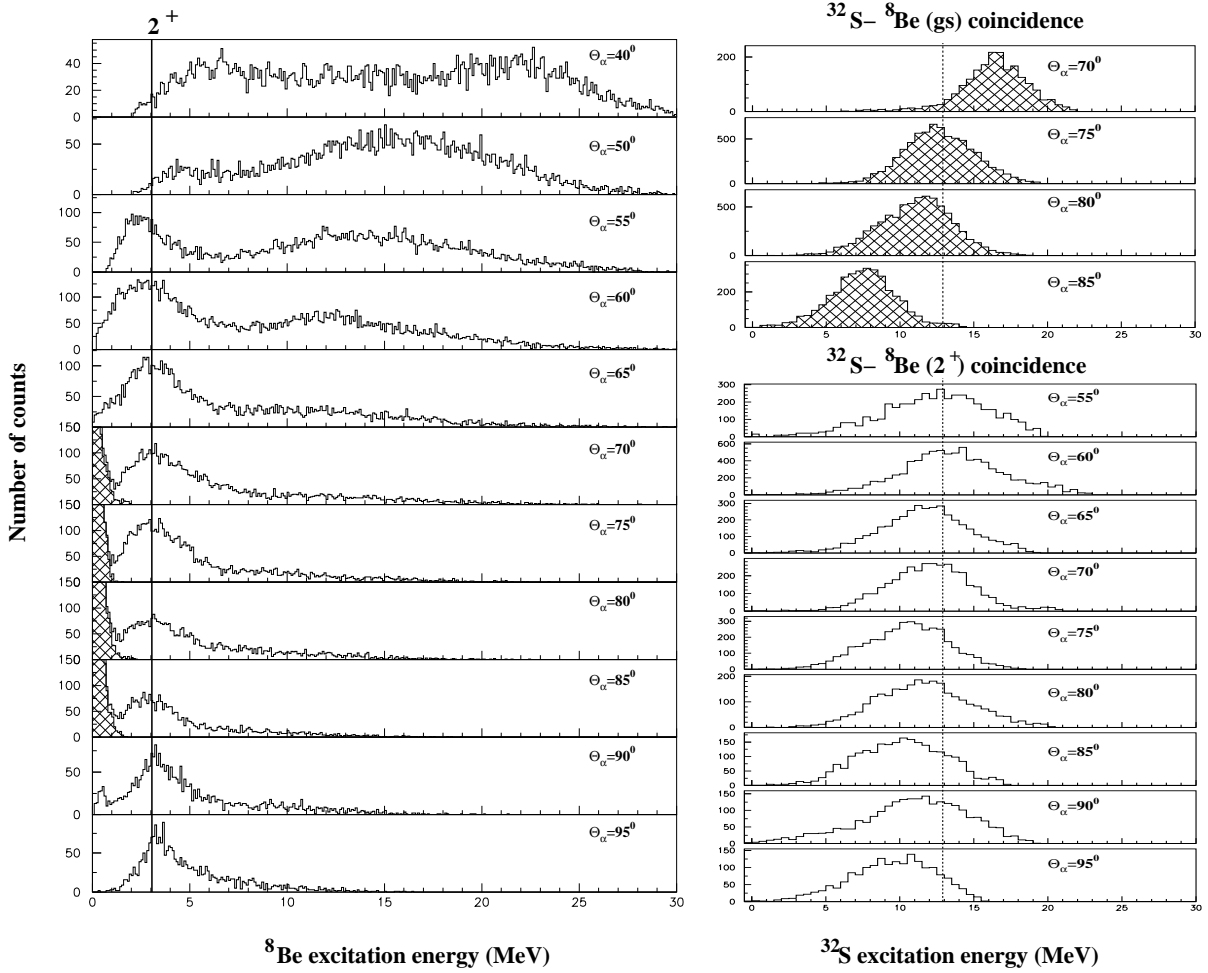}
        \parbox{140mm}{\caption
                      {\label{figure8}  
                      {\it  Calculated excitation energy spectra for  
		      $^{8}$Be (left) and $^{32}$S (right) in coincidence with 
		      the g.s. (up) and first excited level (down) of $^{8}$Be. 
		      The solid line corresponds to the energy of the first 
		      excited state of $^{8}$Be (3.06 MeV). the dashed lines 
		      correspond to $^{32}$S excitation energy expected for an 
		      $\alpha$-transfer process.}}} 
        \end{center}
        \end{figure}

To summarize, the properties of the emitted LCP's in the $^{28}$Si +
$^{12}$C,$^{28}$Si reactions, which are reasonably well described by
statistical-model calculations using spin-dependent level densities, suggest
significant deformation effects at high spin for the $^{40}$Ca and $^{56}$Ni
dinuclear systems. In the case of $^{28}$Si + $^{12}$C the $\alpha$-particle
energy spectra measured in coincidence with a S residue exhibits an additional
component arising from the cluster decay of the unbound $^8$Be nucleus,
produced through the $\alpha$-cluster-transfer of $^{40}$Ca $\rightarrow$
$^{32}$S+$^{8}$Be. This type of reaction allows to populate some N=Z nuclei
with a well defined excitation energy. Thus, sophisticated particle-$\gamma$
experiments (see Refs. [25-28] for instance) using EUROBALL IV and/or
GAMMASPHERE should be performed in the very near future in order to well define
and understand what are the best types of reaction which are capable to
populate significantly the superdeformed bands discovered and/or predicted in
this mass region. 

\vskip 0.3cm

\noindent
{\bf Acknowledgements :} This paper is based upon the Ph.D.~Thesis of Marc
Rousseau, Universit\'e Louis Pasteur, Strasbourg, 2000. We would like to thank
the staff of the VIVITRON for the excellence support in carrying out these
experiments. M.R. would like to acknowledge the Conseil R\'egional d'Alsace for
the financial support of his Ph.D.~Thesis work. Parts of this work has also
been done in collaboration with C.E. during his summer stay at IReS with a
JANUS Grant of IN2P3. 

\vskip 1.0cm

\leftline{\bf References :}

\vskip 0.4cm

\noindent
[1] S.J.~Sanders, A.~Szanto de Toledo, and C.~Beck,   \rm
Phys. Rep. \bf 311\rm, 487 (1999).\\ 
\noindent
[2] S.J.~Sanders,  \rm Phys. Rev. C \bf 44\rm, 2676 (1991).\\ 
\noindent
[3] T.~Matsuse {\it et al.},  \rm Phys. Rev. C {\bf 55}, 1380
(1997).\\ 
\noindent
[4] B.~Shivakumar {\it et al.},  \rm Phys. Rev. C \bf35\rm,
1730 (1987).\\ 
\noindent
[5] D.~Shapira {\it et al.}, \rm Phys. Lett. \bf 114B\rm,
111(1982);  {\it ibidem}, Phys. Rev. Lett. \bf 53\rm, 1634 (1984)\\ 
\noindent
[6] C.E.~Svensson {\it et al.},   \rm Phys. Rev. Lett. \bf
85\rm, 2693 (2000); R.V.F.~Janssens, private communication\\ 
\noindent
[7] G.~Viesti {\it et al.}, Phys. Rev. C {\bf38}, 2640 (1988)
and references therein.\\ 
\noindent
[8] I.M.~Govil {\it et al.},  \rm Phys. Lett. B \bf 197\rm, 515
(1987).\\ 
\noindent
[9] G.~La Rana {\it et al.} Phys. Rev. C {\bf 37}, 1920 (1988);
{\it ibidem} C {\bf 40}, 2425 (1989).\\ 
\noindent
[10] J.~R. Huizenga {\it et al.}, \rm Phys. Rev. C \bf 40\rm,
668 (1989).\\ 
\noindent
[11] I.M.~Govil {\it et al.}, \rm Nucl. Phys. \bf A674\rm, 377
(2000); {\it ibidem}, Phys. Rev. C {\bf 57}, 1269 (1998); {\it ibidem}, Phys.
Lett. {\bf B307}, 283 (1993).\\ 
\noindent
[12] M.~Rousseau, Ph.D. Thesis, Universit\'e Louis Pasteur,
Strasbourg (2000); IReS Report \bf 01-02\rm.\\ 
\noindent
[13] C.~Beck {\it et al.}, Ricerca Scientifica ed Educazione
Permanente (Supp.) \bf 115\rm, 407 (2000).\\ 
\noindent
[14] M.~Kildir {\it et al.}, \rm Phys. Rev. C {\bf 51}, 1873
(1995).\\ 
\noindent
[15] S.K.~Hui {\it et al.}, \rm Nucl. Phys. \bf A641\rm, 21 (1998);
Erratum \bf A645\rm, 605 (1999).\\ 
\noindent
[16] M.F.~Vineyard {\it et al.},  \rm Phys. Rev. C {\bf 47},
2374 (1993).\\ 
\noindent
[17] B.~Fornal {\it et al.}, \rm Phys. Rev. C {\bf 44}, 2588
(1991).\\ 
\noindent
[18] M.F.~Vineyard {\it et al.},  \rm Phys. Rev. C \bf 41\rm,
1005 (1990).\\ 
\noindent
[19] C.~Bhattacharya {\it et al.},  \rm Nucl. Phys. \bf
A654\rm, 841c (1999); {\it ibidem}, Phys. Rev. C (in preparation).\\ 
\noindent
[20] M.~Rousseau {\it et al.},  Clustering Aspects of Nuclear
Structure and Dynamics, eds. M. Korolija, Z. Basrak and R. Caplar (World
Scientific Publishing Co., Singapore, 2000), p.189.\\ 
\noindent
[21] C.~Bhattacharya {\it et al.}, Pramana (Indian Journal
of Physics), to be published (2001); IReS Report {\bf 00-19}.\\ 
\noindent
[22] D.~Mahboub, Ph.D. Thesis, Universit\'e Louis Pasteur,
Strasbourg (1996); CRN Report \bf 96-36\rm.\\ 
\noindent
[23] H.~Morgenstern {\it et al.},  \rm Z. Phys. A \bf 324\rm,
443 (1986).\\ 
\noindent
[24] N.~Arena {\it et al.},   \rm Phys. Rev. C \bf 50\rm, 880
(1994).\\ 
\noindent
[25] R.~Nouicer {\it et al.}, Phys. Rev. C {\bf 60}, 41303
(1999).\\ 
\noindent
[26] C.~Beck {\it et al.}, Phys. Rev. C {\bf 63}, 014607 (2001).\\ 
\noindent
[27] S.~Thummerer {\it et al.}, Phys. Scr. \bf T88\rm, (2000).\\ 
\noindent
[28] S.~Thummerer {\it et al.}, Jour. of Phys. (London) \bf
G\rm (2001), to be published.

\end{document}